\documentclass[11pt]{article}

\usepackage{latexsym,bm,graphicx,color,xcolor,nicefrac,titletoc,enumerate,amsmath,amssymb,xfrac,xcolor,physics,cite,setspace}

\usepackage{mlmodern}
\usepackage[T1]{fontenc}

\usepackage[nottoc]{tocbibind} % add references to TOC

\usepackage{hyperref}
\hypersetup{linktocpage=true,colorlinks=true,linkcolor=blue,citecolor=blue,urlcolor=blue}

\usepackage[skip=3pt plus1pt, indent=20pt]{parskip}

\usepackage[bottom]{footmisc}

\usepackage{geometry}
\usepackage{titlesec}
\titleformat{\section}{\large\bfseries\sffamily}{\thesection}{0.5em}{}
\titleformat{\subsection}{\normalfont\bfseries\sffamily}{\thesubsection}{0.5em}{}
\titleformat{\subsubsection}{\normalsize\itshape\sffamily}{\thesubsubsection}{0.5em}{}
\titleformat*{\paragraph}{\normalsize\bfseries\sffamily}

\numberwithin{equation}{section}

\def\a{\alpha}
\def\b{\beta}

\def\G{\Gamma}
\def\d{\delta}

\def\vf{\varphi}

\def\k{\kappa}

\def\L{\Lambda}
\def\m{\mu}
\def\n{\nu}
\def\r{\rho}
\def\s{\sigma}

\def\pd{\partial}
\def\nab{\nabla}
\def\pr{\prime}

\def\qq{\quad\quad}
\def\q{\quad}

\newcommand{\cG}{\mathcal{G}}

\newcommand{\cL}{\mathcal{L}}

\newcommand{\cO}{\mathcal{O}}

\newcommand{\nfrac}{\nicefrac}

\newcommand{\Ired}{I_{\text{red}}}
\newcommand{\Lred}{L_{\text{red}}}

\newcommand{\link}[1]{[\href{http://arxiv.org/abs/#1}{{\tt arXiv:#1}}]}

\newcommand{\mail}[1]{\href{mailto:#1}{{\tt #1}}}

\begin{document}
	
%\begin{titlepage}       
%	\vspace{5pt} \hfill 
		
	\begin{center}
		%\setstretch{2}
		{\Large \bf \sffamily AdS$\mathbf{_3}$ black holes with primary Proca hair from\\ a regularized Gauss-Bonnet coupling}
	\end{center}
		
	\begin{center}
		\vspace{10pt}
			
		{{\bf \sffamily G{\"o}khan Alka\c{c},}${}^{a}\,${\bf \sffamily  Murat Mesta}${}^{b}\,$ {\bf \sffamily and G{\"o}n\"{u}l \"{U}nal}${}^{c}$}
		\\[4mm]
			
		{\small 
		{\it ${}^a$Department of Aerospace Engineering, Faculty of Engineering,\\ At{\i}l{\i}m University, 06836 Ankara, T\"{u}rkiye}\\[2mm]
				
		{\it ${}^b$Department of Electrical and Electronics Engineering, Faculty of Engineering,\\ At{\i}l{\i}m University, 06836 Ankara, T\"{u}rkiye}\\[2mm]
		
		{\it ${}^c$Program of Biomedical Engineering, Faculty of Engineering,\\ Başkent University, 06790 Ankara, T\"{u}rkiye}\\[2mm]
				
		{\it E-mail:} {\mail{alkac@mail.com}, \mail{murat.mesta@atilim.edu.tr}, \mail{gunalmesta@baskent.edu.tr}}
		}
		\vspace{2mm}
		\end{center}
		
		\centerline{{\bf \sffamily Abstract}}
		\vspace*{1mm}
		\noindent We construct a consistent three-dimensional Einstein-Gauss-Bonnet theory as a vector-tensor theory within the generalized Proca class by employing a regularization procedure based on the Weyl geometry, which was introduced recently in \link{2504.13084}. We then obtain an asymptotically AdS$_3$, static, and circularly symmetric black hole solution with primary Proca hair. Afterward, we investigate the effect of the scalar-tensor Gauss-Bonnet coupling constructed previously by different regularization schemes. We further generalize these solutions by incorporating an electric charge. As special cases, we find a regular black hole solution in addition to charged and uncharged stealth BTZ black hole solutions.
%		\vspace{3mm}
		\par\noindent\rule{\textwidth}{0.5pt}
		\tableofcontents
		\par\noindent\rule{\textwidth}{0.5pt}
%		\newpage
%		\pagestyle{empty}
%\end{titlepage}

%\Large\boldmath\bfseries\
\section{Introduction}
If we assume that field equations of our theories follow from the extremum of an action, we end up with a severe restriction on possible fundamental theories as summarized in the Ostragadski theorem. It states that when the Lagrangian non-degenerately depends on higher-derivatives of dynamical variables, the Hamiltonian of the theory is unbounded from below. Such theories are said to have Ostragadski instabilities since the perturbations grow without bound (see \cite{Woodard:2015zca} for a review). Since this implies that the Lagrangian should contain at most first derivatives to avoid instabilities, it also provides an explanation to why the laws of physics are expressed in terms of second-order differential equations.

One possible workaround for higher-derivatives in a Lagrangian is to have a trivial degeneracy, which is realized when the Lagrangian in question differs from another Lagrangian with at most first derivatives, only by a total derivative. General relativity (GR) is an example scenario where the Einstein-Hilbert action can be written in the following form
\begin{equation}
	\sqrt{-g} R = \sqrt{-g} g^{\m\n} \left[\Gamma^\r_{\, \m \s} \Gamma^\s_{\, \n \r} - \Gamma^\r_{\, \s \r} \Gamma^\s_{\, \m\n}\right] + \partial_\m \left[\sqrt{-g} \left(g^{\n\r} \Gamma^\m_{\, \n\r} - g^{\m\n} \Gamma^\r_{\, \n\r}\right)\right],
\end{equation}
which is the so-called $\Gamma\Gamma$ form of the Lagrangian of GR that he used in his Hamiltonian formulation \cite{Dirac:1958sc}. Although the covariance is lost, one clearly understands why the resulting field equations, the Einstein equations, are second-order differential equations.

One possible generalization of this can be constructed by considering higher-curvature terms in the action, resulting in the Lovelock Lagrangians \cite{Lovelock:1971yv,Lovelock:1972vz,Padmanabhan:2013xyr} given by
\begin{equation}
	\cL_n = \frac{1}{2^n} \delta_{\r_1 \s_1 \ldots \r_n \s_n}^{\m_1 \n_1 \ldots \m_n \n_n} R_{\ \ \ \ \m_1 \n_1}^{\r_1 \s_1} \cdots R_{\ \ \ \ \m_n \n_n}^{\r_n \s_n},
\end{equation}
where $n$ is the order of the Lovelock Lagrangian and $n=1$ corresponds to the Einstein-Hilbert case. The somewhat disappointing part of the story is that the $n$-th order Lovelock Lagrangian contributes to the dynamics only when\footnote{For $d<2n$, $\cL_n=0$ due to full anti-symmetry of the generalized Kronecker delta. $d=2n$ is the critical dimension where $\cL_n$ is a topological invariant.} $d>2n$, which leaves the Einstein-Hilbert action as the only possibility in $d=4$. In order to get a non-trivial contribution to field equations from the leading correction, i.e., the Gauss-Bonnet invariant which is given by
\begin{equation}
	\cL_{n=2} = \cG = R^2 - 4 R_{\m\n} R^{\m\n}  + R_{\m\n\r\s} R^{\m\n\r\s},
\end{equation}
one should work in $d \geq 5$. Although it does not provide a generalization of GR  in $d=4$, Einstein-Gauss-Bonnet gravity defined with the Lagrangian $\cL = R + \a\, \cG$, where $\a$ is a constant, provides an excellent opportunity to test quite a few theoretical ideas beyond GR, especially in the context of AdS/CFT correspondence \cite{Maldacena:1997re,Witten:1998qj,Gubser:1998bc}. Most notably, the holographic studies of the theory in $d \geq 5$ have yielded important results about the micro-causality and the hydrodynamic properties of the dual conformal field theories defined in $(d-1)$-dimensions \cite{Brigante:2008gz,Brigante:2007nu,Buchel:2009sk,Camanho:2009vw,deBoer:2009pn,deBoer:2009gx,Camanho:2009hu}. 

If one does not insist on working with a pure gravity theory, one can construct scalar-tensor and vector-tensor theories in $d<5$ that are also free of Ostragadski instabilities, which are called Horndeski theories \cite{Horndeski:1974wa, Deffayet:2009mn} and generalized Proca theories \cite{Heisenberg:2014rta} respectively. Having second-order field equations, all these theories are natural candidates for admitting exact black hole solutions. As will be explained soon, through different regularization procedures, one can obtain examples of Horndeski or generalized Proca theories starting from Lovelock Lagrangians in $d\geq5$.

After it was realized that GR with a negative cosmological constant admits an AdS$_3$ black hole solution, the Banados-Teitelboim-Zanelli (BTZ) black hole \cite{Banados:1992wn, Banados1993}, gravity in three dimensions has become an active field of research since it allows us to study black hole physics in relatively simple toy models. In particular, we can probe the properties of quantum gravity theories defined in AdS$_3$ spacetime through the AdS$_3$/CFT$_2$ correspondence (see \cite{Kraus:2006wn} for a review) and the infinite number of symmetries present in a 2d CFT \cite{Belavin:1984vu} is a particularly powerful tool in this regard. For example, we can give a microscopic derivation of the semi-classical entropy of the BTZ black hole by using the Cardy formula for the asymptotic growth of the number of states in a 2d CFT \cite{Strominger:1997eq}. Checking whether this can be achieved beyond GR, with solutions different than the BTZ black hole if possible, is a non-trivial test for the generality of the AdS$_3$/CFT$_2$ correspondence. Therefore, it is of importance to obtain AdS$_3$ black hole solutions in different setups including various matter couplings and higher-curvature extensions.

Recently, Glavan and Lin proposed a procedure to define the Einstein-Gauss-Bonnet theory in $d<5$ \cite{Glavan:2019inb}. Studying the solutions of the theory in $d$ dimensions by using an appropriate ansatz, one finds that the contribution of the Gauss-Bonnet term to field equations comes with a factor of $(d-3)(d-4)$. If one scales the coupling constant as $\a \to \frac{\a}{d-p}$ ($p=3, 4$) and then sets $d=p$, it is possible to get novel solutions in $d=3, 4$. While it works for constant curvature spacetimes, cosmological spacetimes and spherically symmetric black holes, it turns out that the field equations for a general metric contain a part which is always  higher-dimensional \cite{Gurses:2020ofy}. Therefore, one does not get a consistent Einstein-Gauss-Bonnet theory in $d=3, 4$ through this procedure.

Although this initial attempt failed, well-defined regularization schemes were discovered \cite{Fernandes:2020nbq,Lu:2020iav,Kobayashi:2020wqy,Hennigar:2020lsl}, which was also applied to the third-order Lovelock Lagrangian \cite{Alkac:2022fuc}.
In order to obtain a consistent lower-dimensional theory, one can either use the difference of the Gauss-Bonnet actions corresponding to two conformally related metrics or perform a Kaluza-Klein reduction with an internal space conformally related to a maximally symmetric space. By again scaling the coupling and then fixing the dimension at the end, not surprisingly, one finds a scalar-tensor theory withing the Horndeski class. For details of the properties of the resulting theories, the reader is referred to \cite{Ma:2020ufk,Hennigar:2020fkv,Hennigar:2020drx,Fernandes:2021dsb,Khodabakhshi:2022knu,Mao:2022zrf,Alkac:2022zda,Bakopoulos:2022gdv,Fernandes:2022zrq,Guajardo:2023uix, Alkac:2023mvr,Babichev:2024krm}. The regularized Lagrangian at the general $n$-th order was obtained in \cite{Fernandes:2025fnz}. By considering an infinite tower of corrections to the Einstein-Hilbert action in $d=4$, which leads to non-local Horndeski theories, it is possible to obtain cosmological models with no big bang singularity, and also regular black hole solutions. Thermodynamics of black hole solutions of these theories was studied in \cite{Cisterna:2025vxk}. Regular black hole solutions in $d=3$ were given in \cite{Fernandes:2025eoc}.

When the maximally symmetric space is taken to be flat, the two methods agree and give the following regularized Gauss-Bonnet invariant in $d=3$
\begin{equation}\label{st}
	\cG_{d\to3}^{\text{st}} = 4 G^{\mu \nu} \varphi_\mu \varphi_\nu-4 X \square \varphi+2 X^2,
\end{equation}
where $\varphi_\mu \equiv \partial_\mu \varphi, \varphi_{\mu \nu} \equiv \nabla_\mu \nabla_\nu \varphi$ and $X \equiv \partial_\mu \varphi \partial^\mu \varphi$. With this at hand, we can define the regularized Einstein-Gauss-Bonnet theory in $d=3$ by the following action
\begin{equation}
	I = \frac{1}{16 \pi G} \int \dd^3{x} \sqrt{-g} \left[R - 2 \L_0 + \alpha \, \cG_{d\to3}^{\text{st}}\right],
\end{equation}
where $\L_0$ is a bare cosmological constant. Looking for a static, circularly symmetric solution of the following form
\begin{equation}\label{anssca}
	\dd{s}^2 = -f(r) \dd{t}^2 + \frac{\dd{r}^2}{f(r)} + r^2 \dd{\theta}^2, \qquad \qquad \vf = \vf(r),
\end{equation}
one finds \cite{Hennigar:2020fkv}
\begin{equation}\label{scasol}
	f = - \frac{r^2}{2 \a} \left(1\pm \sqrt{1-4\a \L_0 + \frac{4 \a m}{r^2}}\,\right), \qquad \qquad \vf = \log (r/r_0).
\end{equation}
The sign in front of the square root can be fixed to be minus by demanding a well-defined $\a \to 0$ limit. Remarkably, the metric function $f$ takes the same form as one gets from the procedure of Glavan and Lin. However, it is now realized as a solution of a particular 3d Horndeski theory and a non-trivial scalar profile supports the solution. Note that there appears no integration constant in the metric function due to the scalar hair in addition the constant $m$, which is related to the mass of the black hole (see \cite{Hennigar:2020fkv} for details) and the geometry is modified only by the coupling constant $\alpha$. Therefore, it is a black hole with secondary hair.

For black hole solutions with a scalar hair, a microscopic derivation of the semi-classical entropy is still possible. However, the Cardy formula must first be written in terms of the ground state energy instead of the central charge of the theory. If the ground state of the dual 2d CFT is identified with the soliton solution, which is obtained from the static black hole solution by a double Wick rotation, the semi-classical entropy can be reproduced \cite{Correa:2010hf, Correa:2011dt, Correa:2012rc}. This idea was successfully applied to the solution \eqref{scasol} of the regularized Einstein-Gauss-Bonnet theory in $d=3$. The semi-classical entropy obeys the area law ($S=\frac{A}{4 G}$ where $A$ is the area of the event horizon) and it was shown to follow from the alternative Cardy formula of \cite{Correa:2010hf, Correa:2011dt, Correa:2012rc} in \cite{Alkac:2024hvu}. 3d Cubic Lovelock gravity obtained in \cite{Alkac:2022fuc} is a more interesting setup. In addition to a black hole solution whose entropy obeys the area law, the theory also admits the BTZ black hole as a solution with a deviation from the area law. In both cases, the alternative Cardy formula works if the corresponding soliton solutions are identified with distinct ground states \cite{Alkac:2024hvu}. This shows that these two black hole solutions belong to disconnected sectors of the theory and each sector has its own ground state.

An arguably more interesting scenario would be to have a black hole solution with primary hair, where an additional integration constant emerges in the metric function. In such a case, considerable deviations from the solutions of GR can be obtained even when the coupling constant is small. However, such solutions are quite rare in the literature. Very recently, by introducing another regularization scheme for the Gauss-Bonnet term within the framework of Weyl geometry, a 4d vector-tensor theory of the generalized Proca class was obtained \cite{Charmousis:2025jpx}. This theory admits asymptotically flat, static, and spherically symmetric black hole solutions with primary hair. For the study of particle motion, shadows, and grey-body factors one may refer to \cite{Lutfuoglu:2025ldc}. 

In this work, motivated by the success of the scalar-tensor 3d Einstein-Gauss-Bonnet theory (and its cubic extension) in testing the AdS$_3$/CFT$_2$ correspondence in a highly non-trivial setup, we construct the vector-tensor 3d Einstein-Gauss-Bonnet theory by using the regularization of \cite{Charmousis:2025jpx}. Studying static and circularly symmetric black hole solutions, we find asymptotically AdS$_3$ solutions with primary hair analogous to those given in \cite{Charmousis:2025jpx}. With these solutions, it becomes possible to check the validity of the prescription of \cite{Correa:2010hf, Correa:2011dt, Correa:2012rc} for black holes with primary hair, which we leave as future work.

The outline of this paper is as follows: In Section \ref{sec:wgb}, we review the procedure of \cite{Charmousis:2025jpx} for the regularization of the Gauss-Bonnet term and present the resulting vector-tensor Lagrangian in $d=3$. In Section \ref{sec:bh}, we derive a black hole solution with primary hair by using this Lagrangian as a correction to the Einstein-Hilbert action. Then, we generalize this solution to a case where the effect of the scalar-tensor Gauss-Bonnet term in \eqref{st} is taken into account. We also find the electrically charged version of the most general solution and we end our paper with a summary of our results followed by some directions for future research in Section \ref{sec:disc}.

%%%%%%%%%%%%%%%%%%%%%%%%%%%%%%%%%%%%%%%%%%%%%%%%%%%%%%%%%%

\section{3d regularized Gauss-Bonnet invariant}\label{sec:wgb}
The logic behind the regularization of \cite{Charmousis:2025jpx} is as follows: The Gauss-Bonnet invariant does not contribute to the dynamics when $d<5$ independent of the connection that is used. Therefore, when we calculate the difference between the Gauss-Bonnet term with the Weyl connection and with the Levi-Civita connection, the difference cannot contribute to the dynamics either, and therefore can only contain terms with $(d-3) (d-4)$  up to boundary terms. Then, a nontrivial Lagrangian contributing to field equations in lower dimensions can be obtained by scaling the coupling and then fixing the dimension at the end as before.

For a torsionless Weyl connection $\widetilde{\G}^\a_{\ \mu \nu}$ with the corresponding covariant derivative $\widetilde{\nab}_\m$, the covariant derivative of the metric does not vanish but takes the following form
\begin{equation}
	\widetilde{\nab}_\a g_{\m\n} = - 2 g_{\m\n} W_\a,
\end{equation}
where $W_\a$ is a vector field. From this, the Weyl connection can be found as
\begin{equation}
\widetilde{\Gamma}^\a{ }_{\mu \nu}=\Gamma^\a_{\ \mu \nu}+\delta^\a_{\ \mu} W_\nu+\delta^\a_{\ \nu} W_\mu-g_{\mu \nu} W^\a,
\end{equation}
where $\Gamma^\a_{\ \mu \nu}$ is the Levi-Civita connection with the covariant derivative $\nab_\m$ satisfying the metric compatibility condition $\nab_\a g_{\m\n} = 0$. After calculating the Riemann tensor for the Weyl connection, all the curvature invariants can be found by contracting the indices. Details can be found in \cite{Dengiz:2011ig, Tanhayi:2012nn, BeltranJimenez:2014iie}. 

An exact black hole solution of 5d Einstein-Weyl-Gauss-Bonnet theory was given in \cite{Bahamonde:2025qtc}. Despite the appearance of additional integration constants initially, they are all fixed by regularity conditions and the metric depends on only the mass and the coupling constant of the Weyl-Gauss-Bonnet term at the end, and one obtains a black hole solution with secondary hair. As we know from \cite{Charmousis:2025jpx}, the situation is different in lower dimensions.

In \cite{Charmousis:2025jpx}, the Gauss-Bonnet term calculated with the Weyl connection was written in the form that allows a regularization as follows
\begin{equation}
	\widetilde{\mathcal{G}}=\mathcal{G}+(d-3) \nabla_\mu J^\mu+(d-3)(d-4) \cL,
\end{equation}
where
\begin{align}
	J^\mu&=8 G^{\mu \nu} W_\nu+4(d-2)\left[W^\mu\left(W^2+\nabla_\nu W^\nu\right)-W_\nu \nabla^\nu W^\mu\right], \\
	\cL&=4 G^{\mu \nu} W_\mu W_\nu+(d-2)\left[4 W^2 \nabla_\mu W^\mu+(d-1) W^4\right],
\end{align}
with the definitions $W^2\equiv W_\mu W^\mu$ and $W^4\equiv (W^2)^2$. As expected, it is equal to the sum of the Gauss-Bonnet term $\cG$ for the Levi-Civita connection and terms with a factor of $(d-3)(d-4)$ up to boundary terms. Disregarding the boundary terms, the regularized Gauss-Bonnet term is given by
\begin{equation}
	\cG_{d\to p}^{\text{vt}} = \lim_{d \to p} \frac{\widetilde{\cG} - \cG}{d-p}, \qq p = 3,4.
\end{equation}
The result in 4d was given in \cite{Charmousis:2025jpx}. In 3d, it reads 
\begin{equation}
\boxed{\cG_{d\to 3}^\text{vt}=-4 G^{\mu\nu}W_\mu W_\nu-4 W^2 \nabla_\mu W^\mu-2 W^4.}
\end{equation} 
Using this invariant, we can define the 3d regularized Einstein-Gauss-Bonnet theory as a vector-tensor theory of generalized Proca class by the following action
\begin{equation}\label{ActionVT}
I = \frac{1}{16 \pi G} \int \dd^3{x} \sqrt{-g} \left[R - 2 \L_0 + \b \, \cG_{d\to3}^{\text{vt}}\right],
\end{equation}
where $\b$ is the corresponding coupling constant.
%%%%%%%%%%%%%%%%%%%%%%%%%%%%%%%%%%%%%%
\section{Black hole solutions with primary hair}\label{sec:bh} 
In order to derive a static and circularly symmetric black hole solution, we take a line element of the form
\begin{equation}\label{LineEle}
\dd s^2 = -N^2(r)f(r) \dd t^2+\frac{\dd r^2}{f(r)}+r^2\dd\theta^2,
\end{equation} 
and assume that the components of the vector $W$ are as follows
\begin{equation}\label{W=w0+w1}
W=w_0(r)\dd t+w_1(r)\dd r.
\end{equation}
By inserting these into the action \eqref{ActionVT}, we can find the reduced action $\Ired = \int \dd r \Lred(N,f,w_0,w_1,r)$ for this type of field configurations. After integrating by parts a few times, the reduced Lagrangian $\Lred$ takes the following form such that it contains at most the first-derivatives of the unknown functions
\begin{align}\label{Lred}
L_\text{red} =& -\frac{\b r w_0^4}{f^2N^3}
+ \frac{2\b r w_0^2w_1N^\pr}{N^2} + 2\b f^2w_1^2(1+rw_1)N^\pr\nonumber
+ \frac{\b w_0^2 \left[(1+2rw_1)f^\pr+2f(w_1+rw_1^2+rw_1^\pr)\right]}{fN}\nonumber\\ 
&-\frac{N}{2}\left[2r\L_0+f^\pr+2\b fw_1^2(1+2rw_1)f^\pr+2\b f^2w_1^2(2w_1+rw_1^2+2rw_1^\pr)\right].
\end{align}
The equations of motion can be found by extremizing the reduced action $\Ired$. Unlike what we have for the scalar-tensor version of the 3d Einstein-Gauss-Bonnet theory, it is not possible to integrate the equations following from the variation of the action directly. However, as in the 4d case \cite{Charmousis:2025jpx}, a close inspection of the equations shows that we have the following symmetry
\begin{equation}\label{Transformations}
w_0^2\to w_0^2+ 2\k f N (1 + r w_1)+ \k^2 r^2, \qq w_1\to w_1+ \k \frac{r}{N f},
\end{equation} 
where $\k$ is an arbitrary constant. Its infinitesimal version is as follows
\begin{equation}\label{infTransformations}
	\d w_0^2 = 2\k f N (2 + r w_1), \qq \d w_1 = \k \frac{r}{N f},
\end{equation}
where we have neglected the $\k^2$ term in the transformation of $w_0^2$. Under these infinitesimal transformations, the reduced action $\Ired$ changes only by a boundary term given by 
\begin{equation}\label{Boundary}
B=2 \k \b f (1+2 r w_1). 
\end{equation}
For our reduced Lagrangian in \eqref{Lred}, the Noether charge corresponding to the global symmetry \eqref{infTransformations} is given by
\begin{equation}
Q=\frac{\pd \Lred}{\pd w_0^\pr}\d w_0+\frac{\pd \Lred}{\pd w_1^\prime}\delta w_1-B,\label{NT}
\end{equation}
which (up to a constant) reads
\begin{equation}\label{NoetherCharge}
Q=\frac{r^2 w_0^2}{f N^2}-(1+r w_1)^2 f.
\end{equation}
This allows us to solve for  $w_0$ in terms of $w_1$ and $f$. With this, one notices that the equations $\d \Ired / \d w_0=0$ and  $\d \Ired / \d w_1=0$ become equivalent. Also, a certain combination of the equations $\d \Ired / \d w_0=0$ and $\d \Ired / \d f=0$ force $N^\pr = 0$. After choosing $N=1$, the three equations become equivalent and $w_1$ can be solved in terms of $f$. Inserting this expression into the equation $\d \Ired / \d N=0$, we can find the metric function as
\begin{equation}
f_\pm(r)=-(m+q) + \frac{r^2}{2\b}\left(1 \pm \sqrt{1+ 4 \b \L_0 + \frac{4\b q}{r^2}}\,\right),
\end{equation}
where we have taken $Q=m+q$ and $q$ is the additional integration constant which makes the solution a black hole with primary hair. The vector components take the following final form
\begin{equation}\label{veccomp}
w_0^2=\frac{(f+m+q)^2}{4 r^2}-c(m+q-f-c r^2), \qq w_1=-\frac{f+m+q}{2 r f}+\frac{c  r}{f},
\end{equation}
where $c$ is another integration constant that does not modify the geometry. The norm of the vector field is constant on-shell, i.e., $W^2=-2c$. Using the symmetry \eqref{Transformations} with $\k = -c$, one gets the same expressions for the components of the vector field with $c=0$. This explains why this constant has no effect on the geometry.

\begin{figure}
	\centering
	\includegraphics[width=\linewidth]{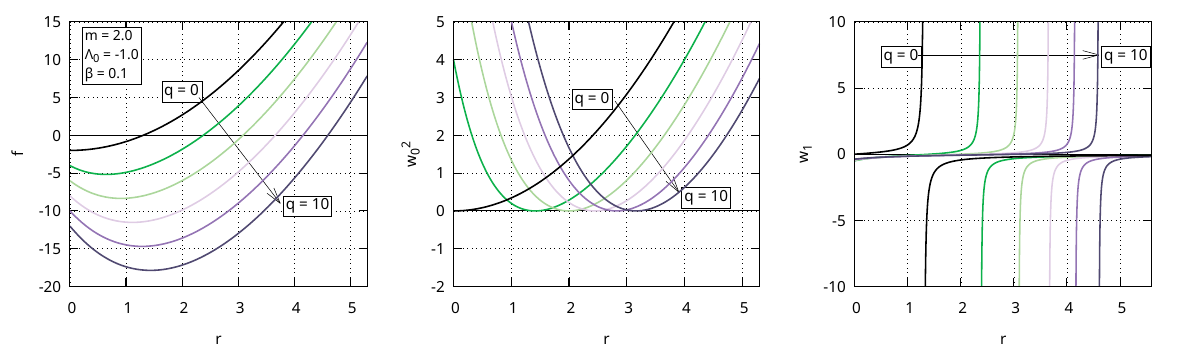}
	\caption{Figure shows behaviors of $f$, $w_0^2$, and $w_1$ for a determined set of parameters $\L_0=-1.0$, $m=2.0$, and $\b=0.1$. Each line is generated for a value of $q\in[0,10]$ as denoted in the figure.}
	\label{fig1}
\end{figure}

The sign in front of the square root can again be fixed by requiring a well-defined limit as the coupling constant goes to zero. As $\b \to 0$, we have
\begin{align}
f_+ &= \frac{r^2}{\b} -m+q+r^2\L_0 - \frac{(q+r^2\L_0)^2}{r^2}\b + \cO(\b^2)\\
f_- &= -m-q-r^2\L_0 + \frac{(q+r^2\L_0)^2}{r^2}\b + \cO(\b^2),
\end{align}
which shows that the $f_-$ branch should be chosen, and from now on, we will proceed with this branch. At large $r$, the metric function behaves as
\begin{equation}
f = - \L r^2 -m-\frac{q}{1+2 \b \L }+ \cO(\nfrac{1}{r^2}),
\end{equation}
where
\begin{equation}
	\L = \frac{-1+\sqrt{1+4 \b \L_0}}{2 \b},
\end{equation}
is the effective cosmological constant. In order to have an asymptotically AdS$_3$ solution ($\L <0$), the following constraints on the parameters should be satisfied
\begin{equation}
	\L_0<0, \qq \qq 0<\b\leq-\frac{1}{4\L_0}.
\end{equation}
As we see, the bare cosmological constant $\Lambda_0$ in the action \eqref{ActionVT} should always be negative and the coupling constant $\b$ should be positive but has an upper bound depending on the bare cosmological constant. In Figure \ref{fig1}, the effect of the hair parameter $q$ is shown for a particular choice of $(m, \L_0, \b)$ satisfying the constraints. One always has a black hole with a single horizon and increasing the value of $q$ shifts the location of the event horizon toward the larger values of $r$. As apparant from \eqref{veccomp}, while $w_0^2$ is finite at the event horizon, $w_1$ diverges.

Note that one gets the metric function of the black hole solution of the scalar-tensor theory in \eqref{scasol} with the identifications $q \to -m$ and $\b \to -\alpha$. In the next section, we will discuss the generalizations of this solution.

\subsection{Effect of the scalar-tensor Gauss-Bonnet term} %%%%%%%%%%%%%%%%%%%%%%%%%%%%%
It is possible to obtain a more general solution by supplementing the action in \eqref{ActionVT} with the scalar-tensor Gauss-Bonnet term given in \eqref{st} as follows
\begin{equation}
	I=\frac{1}{16 \pi G} \int \dd^3x\sqrt{-g} \left (R-2\Lambda_0
	+\a\,\cG_{d\to 3}^{\text{st}}
	+\beta\,\cG_{d\to 3}^{\text{vt}}\right ),
\end{equation}
where $\a$ is the corresponding coupling constant. Equations of motion for the functions $(N, f, w_0, w_1, \vf)$ can again be found from the reduced action. It turns out that a solution is available when we use the profile for the scalar field $\vf$ in \eqref{scasol}, which was also observed in 4d \cite{Charmousis:2025jpx}. 

When $\a \neq \b$, the metric function takes the following form
\begin{equation}
	f = \frac{\b(m+q)}{\a-\b}-\frac{r^2}{2(\a-\b)}\left(1+\sqrt{1-4(\a-\b)\L_0+\frac{4\a\b(m+q)^2}{r^4}+\frac{4\b q}{r^2}}\,\right),
\end{equation}
and the components of the vector $W$ are related to the metric function as in \eqref{veccomp}. As done before, the effective cosmological constant of the theory can be read from the $r \to \infty$ limit as
\begin{equation}
	\L = \frac{1+\sqrt{1-4\L_0(\a-\b)}}{2(\a - \b)}.
\end{equation}
For an asymptotically AdS$_3$ solution, the following constraints should be satisfied
\begin{equation}
	\L_0\geq\frac{1}{4(\a-\b)},\qq \qq \a < \b.
\end{equation}
In this case, one can still have a negative effective cosmological constant despite having a positive bare cosmological constant.

\begin{figure}
    \centering
    \includegraphics[width=\linewidth]{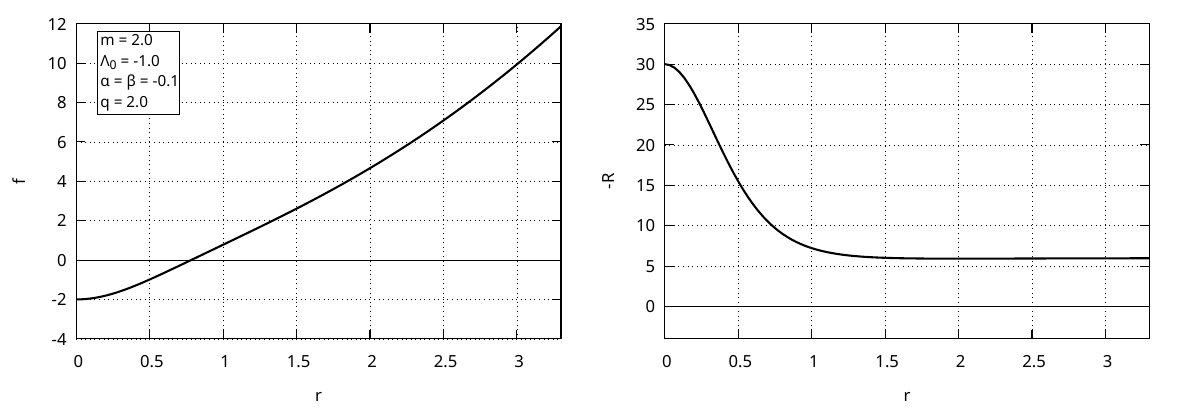}
    \caption{The metric function in \eqref{metric2} and the (minus) Ricci scalar as a function of $r$ are shown for a fixed set of parameters as noted at the top left.}
    \label{fig2}
\end{figure}

When $\a = \b$, one obtains a significantly simpler metric function given by
\begin{equation}\label{metric2}
	f=\frac{\b(m+q)^2+(-\L_0r^2+q)r^2}{r^2-2\b (m+q)},
\end{equation}
with the effective cosmological constant $\L = \L_0$. Studying the Ricci scalar as $r \to 0$, we find
\begin{equation}
	R=\frac{3(m+3q)}{2\b(m+q)} + \cO(r^2).
\end{equation}
In 3d, there are only two more curvature invariants ($R^\mu_{\ \nu} R^\nu_{\ \mu}$ and $R^\mu_{\ \r} R^\r_{\ \nu} R^\n_{\ \mu}$) that can be constructed from the contractions of the metric and the Riemann tensor \cite{Paulos2010, Bueno2022, Gurses2012} and they are also finite as $r \to 0$. If any singularity at positive values of $r$ is avoided by demanding $\b (m+q) <0$, we obtain a regular black hole solution. This puts the following restriction on the parameters
\begin{equation}
	( \b > 0, m < -q)\q \text{ or }\q (\b < 0, m > -q).
\end{equation}
The metric function and the Ricci scalar for a set of parameters satisfying the second constraint can be seen in Figure \ref{fig2}, which shows that the solution is a regular black hole.

Another interesting property of this solution is that one obtains a stealth Banados-Teitelboim-Zanelli (BTZ) black hole solution \cite{Banados:1992wn} for the value of the hair parameter $q=-m$.

\subsection{Addition of an electric charge} %%%%%%%%%%%%%%%%%%%%%%%%%%%%%
We can also find the charged generalizations of black holes presented so far by considering the action
\begin{equation}
	I=\frac{1}{16 \pi G} \int \dd^3x\sqrt{-g} \left (R-2\Lambda_0
	+\a\,\cG_{d\to 3}^{\text{st}}
	+\beta\,\cG_{d\to 3}^{\text{vt}}
	-F_{\m\n} F^{\m\n}\right)
\end{equation}
where $F_{\m\n}=\pd_\m A_\n-\pd_\n A_\m$ is the field strength tensor of the $U(1)$ gauge field $A_\m$. Assuming the following form for the gauge field
\begin{equation}
	A_\m \dd x^\m= \phi(r) \dd t,
\end{equation} 
where $\phi$ is the electric potential, we directly find the Coulomb potential
\begin{equation}
	\phi= -q_e \log(r/r_0),
\end{equation} 
with an electric charge $q_e$. By again using the profile for the scalar field $\vf$ in \eqref{scasol}, we find
\begin{equation}\label{metricQe}
	f = \frac{\b(m+q)}{\a-\b} -\frac{r^2}{2(\a-\b)} \left(1+\sqrt{1-4(\a-\b)\L_0+\frac{4\a\b(m+q)^2}{r^4}+\frac{4\b q-8q_e^2(\a-\b)\log(r/r_0)}{r^2}}\,\right),
\end{equation}
for $\a \neq \b$. The vector components in terms of the metric function $f$ remain the same as in \eqref{veccomp}.

For $\a=\b$, we get the following much simpler metric function
\begin{equation}
	f=\frac{\b (m+q)^2+(-\L_0r^2+q) r^2-2q_e^2r^2\log(r/r_0)}{r^2-2\b (m+q)}.
\end{equation} 
As expected, it becomes a stealth charged BTZ black hole solution for $q=-m$. One might wonder whether this solution is also regular. Due to the logarithmic behavior of the electric potential, the Ricci scalar is divergent as $r\to0$, that is,
\begin{equation}
	R = \frac{3(m + 3 q) -2 q_e^2 \left[5 + 6 \log(r/r_0)\right]}{2\b (m+q)} + \cO(r^2).
\end{equation}

\section{Summary and outlook}\label{sec:disc} %%%%%%%%%%%%%%%%%%%%%%%%%%%%%
In this paper, we constructed the 3d Einstein-Gauss-Bonnet theory as a vector-tensor theory within the generalized Proca class by using the regularization procedure introduced recently in \cite{Charmousis:2025jpx}. We studied the static and circularly symmetric black hole solutions of this theory. As shown in \cite{Charmousis:2025jpx} for 4d, unlike the solutions of the scalar-tensor theories obtained as lower-dimensional limits by different regularization schemes, the resulting equations of motion for the unknown functions can only be solved with the help of the Noether charge associated with a global symmetry of the reduced action. We found an AdS$_3$ black hole with primary hair, which has only a few examples in the literature \cite{Priyadarshinee:2023cmi,Sardeshpande:2024bnk}. We then generalized it by introducing the scalar-tensor Gauss-Bonnet invariant into the action. When the coupling constants of the scalar-tensor and vector-tensor invariants are equal, we obtained a regular black hole solution. Additionally, we presented the electrically charged version and as a by-product, we observed that stealth BTZ and charged BTZ black hole solutions arise for an appropriate choice of the hair parameter.

Our work opens up quite a few possibilities for future work. First of all, in order to understand the physical significance of the hair parameter $q$, thermodynamics of these solutions should be studied. As in the case of the scalar-tensor 3d Einstein-Gauss-Bonnet theory \cite{Hennigar:2020drx}, rotating black hole solutions can be obtained by  boosting static solutions \cite{Lemos:1994xp, Lemos:1995cm}. The scalar-tensor 3d Einstein-Gauss-Bonnet theory together with its cubic extension constructed in \cite{Alkac:2022fuc} exhibit remarkable holographic properties. As mentioned in the introduction, it is possible to give a microscopic derivation of the entropy of their static black hole solutions \cite{Alkac:2024hvu}. Additionally, for these theories, one can construct a monotonically increasing c-function whose value at the UV fixed point matches with the Weyl anomaly coefficient \cite{Alkac:2022zda}. This means that they admit a holographic c-theorem \cite{Freedman:1999gp,Myers:2010xs,Myers:2010tj}, which mimics the Zamolodchikov's c-theorem for 2d QFTs. Theories admitting a holographic c-theorem are quite rare and the only other 3d examples are new massive gravity \cite{Bergshoeff2009,Sinha:2010ai}, its certain higher-curvature \cite{Paulos2010} and Born-Infeld type \cite{Gullu:2010pc,Gullu:2010st,Alkac:2018whk} extensions. A natural question is whether these properties are shared by the 3d vector-tensor Einstein-Gauss-Bonnet theory that we constructed here and its cubic extension. Since the black hole solutions presented here possess primary hair, the corresponding ground states used in the microscopic derivation of the semi-classical entropy might exhibit particularly interesting properties not observed before. Very recently, an alternative regularization mechanism has been proposed\cite{Eichhorn:2025pgy}, where the difference between Gauss-Bonnet invariants with two different Weyl vectors is used. The resulting bi-Proca theory gives rise to regular black hole solutions. A similar behavior is expected in 3d.

\paragraph*{Note added} After we submitted our manuscript to arXiv, another paper studying black hole solutions of 3d vector-tensor regularized Einstein-Gauss-Bonnet thoery appeared on arXiv \cite{Liu:2025dqg}. The authors also provide the metric function given by us in \eqref{metricQe} corresponding to the most general solution (up to a redefinition of integration constants). However, there are important differences in the approaches used: The authors of \cite{Liu:2025dqg} directly use a particular relation between the vector components $w_0$ and $w_1$ as an ansatz, and choose the condition $N^\pr=0$ after obtaining the Euler-Lagrange equations resulting from the reduced action. This approach gives a solution in which the norm of the vector $W$ vanishes on-shell. On the other hand, we find a global symmetry of the equations of motion without assuming any relation between $w_0$ and $w_1$. We then decouple the equations by making use of the Noether charge that allows us to find out that $N^\pr=0$ is a must, and there is a more general relation between the vector components such that the norm is a constant, which can be shifted by the global symmetry.

\paragraph*{Acknowledgements} G. A. acknowledges funding from the Scientific and Technological Research Council of Turkey (T\"{U}B\.{I}TAK) under Project No. 124F058. We are grateful to Mokhtar Hassaine for helpful discussions and for clarifying certain details of \cite{Charmousis:2025jpx} to us.

\end{document}